\documentclass{article}

\usepackage{graphicx}
\usepackage[french]{babel}
\usepackage[latin1]{inputenc}
\usepackage{subfig}

\usepackage{wrapfig}
\usepackage{multicol}
\usepackage[colorlinks=false]{hyperref}
\usepackage[final]{pdfpages}
\usepackage{array} 

\usepackage{tikz}
\usepackage{pdflscape}
\usepackage{color}

\newcommand{\tex}{\TeX\xspace}

\title{Free software, Open source software, licenses.\\
\large{A short presentation including a procedure for research software 
and data dissemination}}

\author{T. Gomez-Diaz\\
\small{CNRS, Université Paris-Est, Laboratoire d'informatique
Gaspard-Monge}\\
\small{77454 Marne-la-Vallée Cedex 2, France}\\
\small{Contact: \texttt{Teresa.Gomez-Diaz@u-pem.fr}}}

\date{9 september 2014, V.1.2\\
\small{This work is licensed under a Creative Commons Attribution-ShareAlike 4.0
International License\\
(CC-BY-SA v4.0), {\url{http://creativecommons.org/licenses/by-sa/4.0/}}}}

\begin{document}

\maketitle

\section{Introduction}

As H2020 approaches and open access policies spread in the European
research community (\cite{EUComm}),
it becomes more important to revisit the basic concepts in software
distribution such as free software, open source software, licenses:
they should be considered before any research software 
dissemination.
We also propose here a distribution procedure that can be adapted for software and
data dissemination.

The target public of this document is the research community at large.
This document is at your disposal under
the license \href{http://creativecommons.org/licenses/by-sa/4.0/}{CC-BY-SA
v4.0}, which means that it can be freely modified,
but authors of modifications agree to keep the initial author information
and state the changes made.
Redistribution of this document or its modified version should be done under
the same license terms.
If you find this document useful, please make it available to a large public.

\section{Definition of Free software}

Free software has been first defined by the Free Software Foundation (FSF) created in 1985
by Richard M. Stallman, but free software was already available. Among well
known examples there are \tex{} by Donald Knuth (1978) or
the Berkeley Software Distribution (BSD) by the Computer Systems Research
Group of the University of California (1977-1995).

This definition can be found on the page {\url{http://www.gnu.org/philosophy/free-sw.html}}.

\begin{quote}
``A program is \textit{free software} if the program's users have the four essential freedoms:

\begin{description}
\item[Freedom 0] The freedom to run the program as you wish, for any purpose.
\item[Freedom 1] The freedom to \textbf{study} how the program works, and change it so it does 
your computing as you wish. Access to the source code is a precondition for this.
\item[Freedom 2] The freedom to redistribute copies so you can help your neighbor.
\item[Freedom 3] The freedom to distribute copies of your modified versions to others. By doing 
this you can give the whole community a chance to benefit from your changes. 
Access to the source code is a precondition for this."
\end{description}
\end{quote}

A program, code or software is free software because it is distributed respecting these four liberties.
Just to say that a software is free means nothing, unless the above mentioned four liberties are
(legally) insured through the license (called free software license) that comes with the software. 
If one liberty is missing the software cannot be free software, and it can be called
proprietary software.

In our opinion, there still remains in some circles a misunderstanding related
to the fourth liberty, therefore we would like to stress here that the distribution of
modifications remains the choice of their author, 
and once redistribution is decided, it is necessary to verify that the reciprocity clauses of
the initial software license are respected.

Note the importance of the word \textit{study} in this definition,
which was born in circles near research institutions. The FSF community states that only
free software should be used in education.
Our personal belief is that free software is particularly well adapted to the research 
environment and the way science evolves.

\section{Definition of Open source software}

By the end of the 90's, the software community forked and the Open Source Initiative (OSI)
was created in 1998 to define the concept of open source in the form of ten conditions.
The full definition can be found on the page {\url{http://www.opensource.org/docs/osd}}, and
here we only sketch the main items.

\begin{quote}
``Open source doesn't just mean access to the source code. The distribution
terms of \textit{open-source software} must comply with the following criteria:

\begin{enumerate}
\item Free Redistribution

The license shall not restrict any party from selling or giving away the software as a component of 
an aggregate software distribution containing programs from several different sources. 
The license shall not require a royalty or other fee for such sale.

\item Source Code
\item Derived Works
\item Integrity of The Author's Source Code
\item No Discrimination Against Persons or Groups
\item No Discrimination Against Fields of Endeavor
\item Distribution of License
\item License Must Not Be Specific to a Product
\item License Must Not Restrict Other Software
\item License Must Be Technology-Neutral"
\end{enumerate}
\end{quote}

We have included the full first item as it shows that this concept is closest to commercial 
interests rather than to liberty or to study goals.
Again, a software is open source because the (open source) license that comes with the software
respects these ten principles, and if one is missing, it is not open source software.

\section{Differences, terminology, licenses}

These two definitions correspond to (very) different philosophies,
but in fact most software we are talking about here is at the same time
free and open source,
as the most commonly used licenses verify the conditions of both definitions.
In legal terms, free and/or open source licenses are contracts and contribute to 
the legal framework where one is allowed to use, copy, modify, and redistribute the software.
Free or open source software is not ``free of rights", author's rights are not going to disappear
because a software is free or open source. In fact, the licenses extend the initial legal frame
stated by the law (copyright law, \textit{code de la propriété
intellectuelle}, \textit{derechos de autor}...). 

Nevertheless there are examples of software which is open source but not
free, we mention here two kind of examples.
\begin{itemize}
\item Software using license NASA v1.3:
\begin{itemize}
\item this license is listed in the OSI list of licenses ({\url{http://opensource.org/licenses}}), see
{\url{http://www.opensource.org/licenses/nasa1.3}}
\item but it is listed by the FSF in the Nonfree Software Licenses 
({\url{http://www.gnu.org/licenses/license-list.html}}),
see {\url{http://directory.fsf.org/wiki/License:NASA-OSA_v1.3}}
\begin{quote}\textit{
The NASA Open Source Agreement, version 1.3, is not a free software license because 
it includes a provision requiring changes to be your ``original creation" ...}
\end{quote}
\end{itemize}
\item Another example becomes more and more common as the use of software
is nowadays present in phones and other electronic materials. These
devices can include executable 
versions of free software but sometimes do not allow to change the version of the
executable, and so it is not possible to use the free software in a free way\footnote{For more
information you can see 
{\url{https://www.gnu.org/philosophy/open-source-misses-the-point.html}}}.
\end{itemize}

Some more terminology: \textit{commercial software} is software with a license that requires a financial counterpart.
Free and open source software can be commercial, free doesn't always mean \textit{gratis}\footnote{Free can translate
as \textit{gratis} in Spanish or \textit{gratuit} in French, which is very
different from \textit{libre}.}
and the opposite also applies: \textit{gratis} doesn't always correspond
to free or open source software.
Software distributed with source code is not always free or open source. Software distributed
without a license is not free nor open source, it is all rights reserved.

There are also members of the software community who use the terminology of
Free/Open Source Software (FOSS) or Free/Libre/Open Source Software (FLOSS)
(where \textit{libre} remains in French or Spanish to stress the liberty side of the word free)
because they prefer not to choose between the two definitions.

\section{A classification of free/open source licenses}

Licenses give rights (and liberties) but also have reciprocity clauses to be respected, as
for example not to give the initial name to a modified version, or the requirement to cite some work or publication.
Here we give a classification extracted from ``A Practical Guide ..." by
T. Aimé (\cite{Aime}, p.9, see also \cite{FAQ}).

\begin{table}[h]
\captionsetup{font=small}
\centering
\begin{tabular}{l r}
\begin{minipage}{5.5cm}

\includegraphics[scale=0.5,bb=25 230 0 0]{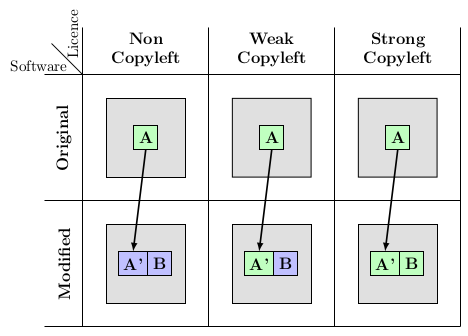}
\vspace{4.2cm}
\hspace{0.4cm}\caption{\small A classification of licenses.}
\end{minipage}
&
\begin{minipage}{10.5cm}
\begin{itemize}
\item [Strong copyleft:] the initial software license of A is inherited in
software A' (without regarding if it has been modified) and new linked
components B. The reciprocity clause indicates that the initial license
remains at the distribution of the modified software. Well known examples
are \href{http://www.gnu.org/copyleft/gpl.html}{GNU GPL} and the French
\href{http://www.cecill.info/licences/Licence_CeCILL_V2.1-en.html}{CeCILL v2}.
\item [Weak copyleft:] the initial software license of A remains in A or its
modified version A', but new software components may have any
license. For example see licenses
\href{http://en.wikipedia.org/wiki/Mozilla_Public_License}{MPL},
\href{http://www.gnu.org/copyleft/lgpl.html}{GNU LGPL},
\href{http://www.cecill.info/licences/Licence_CeCILL-C_V1-en.html}{CeCILL-C}.
\item [Non copyleft:] the initial software license of A can disappear in A'
(without regarding if it has been modified) and new components can have
any kind of license. For example see licenses
\href{http://en.wikipedia.org/wiki/Apache_License}{Apache},
\href{http://en.wikipedia.org/wiki/BSD_License}{BSD},
\href{http://en.wikipedia.org/wiki/MIT_License}{MIT},
\href{http://www.cecill.info/licences/Licence_CeCILL-B_V1-en.html}{CeCILL-B}.
\end{itemize}
\end{minipage}
\end{tabular}
\end{table}

Because of its impact, we quote here 
the GPLv2 phrase that activates license inheritance:
\textit{``You must cause any work that you distribute or
publish, that in whole or in part contains or is derived from
the Program or any part thereof, to be licensed
as a whole at no charge to all third parties under the terms of this
License."}

Beware of possible conflicting obligations imposed by licenses
when including and modifying a lot of software components to build
a new one. You can find here the FSF license compatibility table:
\url{http://www.gnu.org/licenses/gpl-faq.html\#AllCompatibility}.

\section{The double rôle of licenses}

On the one hand, licenses establish the legal framework\footnote{Modification of any
legal information (authors, institutions, dates, licenses...) in components
gathered on the net should be avoided.}
in which it is allowed to use, copy, modify, and redistribute software. In France, law
indicates that it is illegal to use or to modify software unless
you have a (written) agreement to do so (see \cite{Aime}, p.3).
Giving access to software in web pages does not produce free nor open source
software, nor legally
useable software: it is important to establish a license. 
Free and open source licenses give a legal frame ``by default", and if it is not convenient for
the use or support you need, you can contact authors to establish
other kind of collaborations (usually by paying more services).

On the other hand, when the use of licenses such as free software licenses
is decided at the level of head institutions, laboratories or research
funding agencies, when they are required in documents such as the Berlin
declaration (\cite{Berlin})\footnote{See also the Budapest Open Acces
Initiative (\cite{Buda}), where licenses are not mentioned, but are suggested:
``\textit{... permitting any users to read, download, copy, distribute, print,
search, or link, ...}"}, this corresponds to strong policies where free
access to the research production is at stake, having as consequence, for
example, to increase the reproductibility of research.

The word \textit{open} has quickly expanded beyond the software limits, we are
now in the world of open access, open education, open innovation, open
data, ... and even open government.
It has become also a fancy word and it can appear
in contexts not ``open" at all; please be careful and check definitions, licenses, policies.

\section{A distribution procedure for research software or data}

The following list of steps can be adapted in order to create a distribution
procedure for research software (see \cite{Recom}).
It can also be adapted to distribute research data.
Steps marked with (*) are to be revisited regularly in each version release.

\begin{itemize}
\item Choose a name or title to identify the object, avoid trademarks and
other proprietary names, you can associate date, version number,
target platform...
\item (*) Establish the list of authors and affiliations. An associated percentage of participation, completed with 
minor contributors can be useful. If the list is too long, keep updated
information in a web page.
\item (*) Establish the list of main functionalities.
\item (*) Establish the list of included software and data components,
indicate their licenses or other documents giving right to access, copy, modification,
redistribution for the component.
\item Choose a license, with the agreement of all the rights' holders and authors, have a signed
agreement if possible. Consider using free software licenses 
and \href{https://creativecommons.org/choose/}{CC licenses (v4.0)} for
data (for example).
\item Choose a web site, forge, or deposit to distribute your product; licensing and conditions
of use, copy, modification, and/or redistribution should be clearly stated, as well as the best way
to cite your work. Good metadata and respect of open standards are always
important when giving away new components to a large community: it helps
others to use your work and increases its longevity.
Give licenses to the documentation
(\href{http://www.gnu.org/copyleft/fdl.html}{GNU FDL},
\href{https://creativecommons.org/choose/}{CC},
\href{http://artlibre.org/licence/lal/en/}{LAL}...) and to web
sites. Use
\href{http://en.wikipedia.org/wiki/Persistent_identifier}{persistent
identifiers} if possible.
\item (*) In order to help tracking new important functionalities, archive a tar.gz or similar
in safe place.
\item Inform your laboratories and head institutions (if this has not be done in the license step).
\item Create and indicate clearly an adress of contact.
\item Distribute the software or data component.
\item Inform the community.
\end{itemize}

\paragraph{How to give a license.}

To install a license is not a difficult step, but it should be done
before distributing the software. The basic idea is to identify the files
with a commented heading including the important informations about authors,
dates, licenses, and to add to the whole set of files an additional file including
the text of the license (or a link to a web page with it) in a file
named COPYING, LICENCE or README for example.
Please indicate the license in the documentation, in the web site, as well as the authors
and their affiliations.
Here is an example for the headings:
\begin{itemize}
\item name: name of the file, name and version of the software,
\item holder of rights: Copyright (©), year(s), head institutions,
persons...
\item authors: the list of authors and their institutions (or a link if it
is too long),
\item license: name and version,
\item dates: creation date, last modification date.
\end{itemize}

\section{Conclusion}

The main goal of this document is to help the research community to
understand the basic concepts of software distribution and the associated licenses.
If you think that this document is useful in this respect, please help
distributing this document to a wider audience.

\small{\textbf{Acknowledgements.} With many thanks to proofreaders
of previous versions of this document.}

\renewcommand{\refname}{References}

\end{document}